\documentclass[aps,prd,twocolumn,groupedaddress,showpacs]{revtex4-1}

\usepackage{graphicx}
\usepackage{units}
\usepackage[intlimits]{amsmath}
\usepackage{amssymb}
\usepackage{hyperref}
\usepackage{siunitx}
\usepackage{multirow} 

\usepackage{amsfonts}
\usepackage{amsmath}
\usepackage{bm}                 

\newcommand{\elabel}[1]{\label{eqn:#1}}

\newcommand{\cref}[1]{\ref{cpt:#1}}

\newcommand{\eref}[1]{\ref{eqn:#1}}

\newcommand{\listBegin}{\begin{tabular}{cp{4.5in}}}
\newcommand{\listEnd}{\end{tabular}}

\newcommand{\matrixBegin}[1]{\left[\!\!\left[ \begin{array}{#1}}
\newcommand{\matrixEnd}{\end{array} \right]\!\!\right]}

\newcommand{\beq}[1]{\begin{equation}\elabel{#1}}
\newcommand{\eeq}{\end{equation}}

\newcommand{\beqa}[1]{\begin{eqnarray}\elabel{#1}}
\newcommand{\eeqa}{\end{eqnarray}}

\newcommand{\abs}[1]{\left| #1 \right|}

\newcommand{\fun}[2]{\,{#1}\!\left( {#2} \right)}

\newcommand{\tfun}[2]{\fun{\textrm{#1}}{#2}}

\newcommand{\imag}[1]{Im\!\left[ {#1} \right]}
\newcommand{\real}[1]{Re\!\left[ {#1} \right]}

\newcommand{\partderivS}[2]{\frac{\partial^2{#1}}{\partial{#2}^2}}

\usepackage{siunitx}

\DeclareSIUnit\rtHz{$\sqrt{\si{\Hz}}$}
\DeclareSIUnit\mrtHz{\meter\per\rtHz}

\usepackage{color}
\definecolor{spring}{rgb}{0.7,0.9,0.7}
\definecolor{brick}{rgb}{0.7,0.2,0.1}
\definecolor{redHL}{rgb}{1.0,0.5,0.5}

\newcommand{\smallMatrix}[1]{\begin{pmatrix}#1\end{pmatrix}}

\def\I{\mathrm{i}}
\newcommand{\E}[1]{\mathrm{e}^{#1}}

\def\Fhd{\mathbf{\bar{b}}_{\zeta}}

\def\Sqz{\mathbf{S}}  
\def\Eye{\mathbf{I}}   
\def\Rot{\mathbf{R}}  

 \newcommand{\TnN}[1]{\mathbf{T_{#1}}}
\def\Tn{\TnN{n}}
\def\Tinj{\TnN{inj}}
\def\Tfc{\TnN{fc}}

\def\Tifo{\TnN{ifo}}
\def\Tsqz{\TnN{1}}
\def\Tmm{\TnN{mm}}

\def\rfc{r_{\rm fc}}
\def\gfc{{\gamma_{\rm fc}}}
\def\omfc{{\omega_{\rm fc}}}
\def\Osql{{\Omega_{\rm SQL}}}

\def\Kifo{\mathcal{K}}
\DeclareMathOperator{\arccot}{arccot}
\DeclareMathOperator{\arcsinh}{arcsinh}
\def\ifo{{\rm ifo}}
\def\arm{{\rm arm}}

\def\Loss{\Lambda}
\newcommand{\Attn}{\tau}

\def\TwoP{\mathbf{A_2}}

\def\mmSqz{a}
\def\mmLO{b}

\def\Usqz{U_{\rm sqz}}
\def\Ufc{U_0}
\def\Ulo{U_{\rm lo}}
\def\Urfc{U_{\rm rfc}}

\def\dA{\delta \! A}
\def\dAf{\delta \! \tilde A}
\def\dP{\delta \! P}
\def\dPf{\delta \! \tilde P}
\def\sE{\mathcal{E}_0}
\def\sA{\mathcal{A}}

\begin{document}

\title{Decoherence and degradation of squeezed states in quantum filter cavities}

\author{P Kwee}
\author{J Miller}
\email{jmiller@ligo.mit.edu} 
\author{T Isogai}
\author{L Barsotti}
\author{M Evans}

\affiliation{LIGO Laboratory, Massachusetts Institute of Technology, 185
  Albany St, Cambridge, MA 02139, USA} 

\date{\today}


\begin{abstract}
  Squeezed states of light have been successfully employed in
  interferometric gravitational-wave detectors to reduce quantum
  noise, thus becoming one of the most promising options for extending
  the astrophysical reach of the generation of detectors currently
  under construction worldwide. In these advanced instruments, quantum
  noise will limit sensitivity over the entire detection
  band. Therefore, to obtain the greatest benefit from squeezing, the
  injected squeezed state must be filtered using a long-storage-time
  optical resonator, or ``filter cavity'', so as to realise a
  frequency dependent rotation of the squeezed quadrature. Whilst the
  ultimate performance of a filter cavity is determined by its storage
  time, several practical decoherence and degradation mechanisms limit
  the experimentally achievable quantum noise reduction. In this paper
  we develop an analytical model to explore these mechanisms in
  detail. As an example, we apply our results to the \SI{16}{\m}
  filter cavity design currently under consideration for the Advanced
  LIGO interferometers.
  \end{abstract}

\pacs{04.80.Nn, 42.50.Dv, 04.30.-w, 42.50.Lc} 

\maketitle

\section{Introduction}
Squeezed states of light are used in a variety of experiments in
optical communication, biological sensing and precision measurement
\cite{Tak13a,Tay13a,Yon12a}. To gravitational-wave detectors, the
finest position-meters ever built, squeezed states of light today
represent one of the most mature technologies for further expanding
the detectable volume of the universe~\cite{LIG11b, Bar13a}.

The advanced detectors currently under construction, such as Advanced
LIGO \cite{Har10a}, will be limited by quantum noise over their entire
detection band, from 10 Hz to 10 kHz.  To fully exploit the potential
of squeezing, squeezed states must therefore be manipulated so as to
impress a frequency dependent rotation upon the squeezing
ellipse. Such rotation can be realised by reflecting the squeezed
states from a detuned, over-coupled, optical resonator, called a
quantum filter cavity.

The performance of ideal filter cavities, fundamentally limited by
their storage times, is well-understood \cite{Kim01a,Har03a} and a
proof-of-principle experimental demonstration has been performed
\cite{Che05a}. However, the impact of several decoherence and
degradation mechanisms which critically determine the achievable
performance of astrophysically relevant filter cavities has not yet
been investigated.

In this paper we present an analytical model, based on the two-photon
formalism~\cite{Cav85a,Sch85a,Cor05a}, which evaluates the reduction
in observable squeezing caused by optical losses and by spatial mode
mismatch between the injected squeezed light, the filter cavity and
the interferometer.  Further, we also explore the influence of
squeezed quadrature fluctuations \cite{Dwy13}, or ``phase noise'',
generated both inside and outside the filter cavity.  As a concrete
example, we study the effects of these noise sources on a 16 m long
filter cavity with a 60 Hz linewidth, parameters considered for
Advanced LIGO~\cite{Eva13a}.

\section{Analytical model}
\label{sec:model}
\begin{figure*}[t!]
   \includegraphics[width=0.675\textwidth]{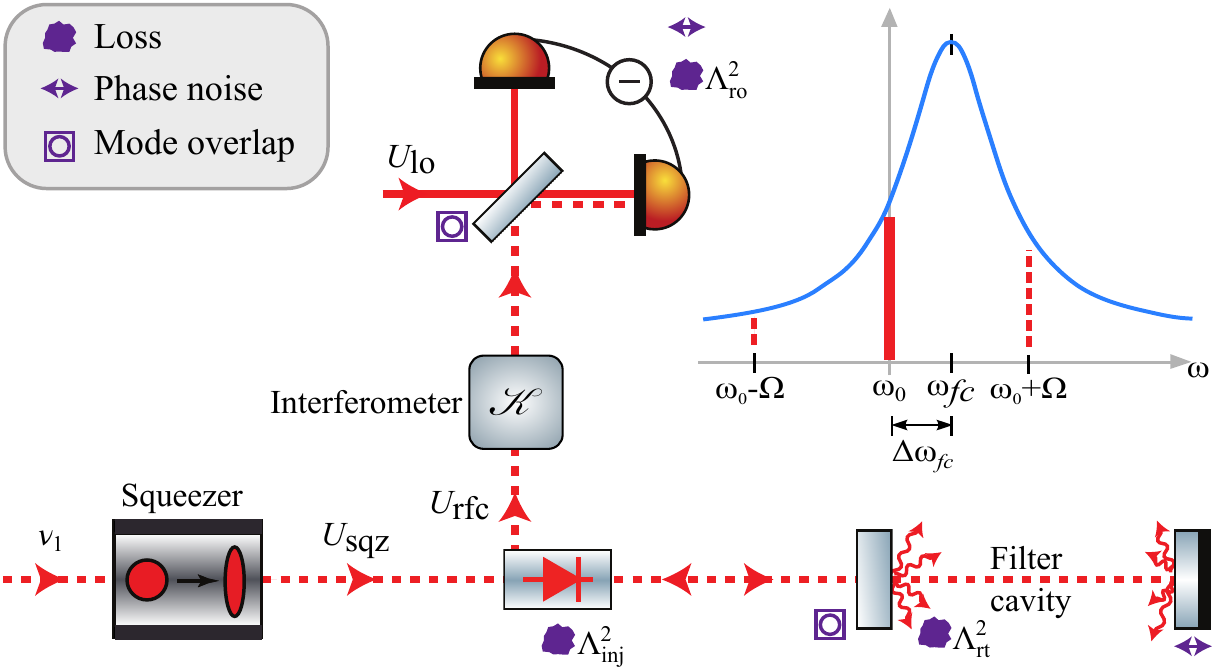}
    \caption{The frequency dependent squeezing system analysed in this
      work.  The squeezer generates a frequency-independent squeezed
      state with spatial mode $\Usqz$. The squeezed state becomes
      frequency dependent after reflection from a filter cavity and is
      subsequently detected via homodyne readout using a local
      oscillator with spatial mode $\Ulo$.\label{fig:model}}
\end{figure*}

The frequency dependent squeezing system modelled in this work is shown
in Figure~\ref{fig:model}. The squeezed beam is injected into the
interferometer after reflection from the filter cavity. In this model
we assume that the quantum noise enhancement is measured via a generic
homodyne readout system, by beating the interferometer output field
against a local oscillator (LO) field. The main sources of squeezing
decoherence (optical loss and mode-mismatch) and degradation (phase
noise due to local-oscillator phase-lock errors and cavity length
fluctuations) are indicated.

Using the mathematical formalism described in~\cite{Eva13a} and
further developed in appendix \ref{app:Formalism}, our analysis
calculates the achievable quantum noise reduction by propagating three
classes of vacuum field through the optical system: $v_1$ which
passes through the squeezer and becomes the squeezed field; $v_2$
which accounts for all vacuum fluctuations that are coupled into the
beam due to optical losses \emph{before} the interferometer; and $v_3$
which accounts for vacuum fluctuations introduced due to losses
\emph{after} the interferometer. In this formalism vacuum fields are
proportional to the identity matrix,
\hbox{$v_1=v_2=v_3=\sqrt{2\hbar\omega_0}\mathbf{I}$}, and their
interaction with an optical element or system may be described by
multiplication with a $2\times2$ transmission matrix $\mathbf{T}$,
i.e.~$v_{\mathrm{out}}=\mathbf{T}v_{\mathrm{in}}$.

In sections \ref{sec:inj}, \ref{sec:filterCavity} and
\ref{sec:modeMatching} we develop transfer matrices for the
propagation of $v_1$ through the squeezer and injection optics, its
modification by the filter cavity and the influence it experiences due
imperfect mode-matching. Section \ref{sec:interferometer}
constructs a transfer matrix describing the optomechanical coupling of
the interferometer and shows that it can be written as a product of
rotation and squeezing operators. We then, in section
\ref{sec:LNT}, incorporate the uncontrolled vacuum noise coupled into
the squeezed field due to loss and show how one can compute the
quantum noise at the readout of the interferometer using the matrices
developed in the previous sections. The final piece of our
analytical model, performance degradation due to phase noise, is
detailed in section \ref{sec:phaseNoise}.

\subsection{Squeezed field injection}
\label{sec:inj}
The squeezer is represented by the operator $\Sqz(\sigma, \phi)$, given by
\begin{align}
\Sqz(\sigma, \phi) &= \Rot(\phi) \Sqz(\sigma,0) \Rot(-\phi) = \Rot_\phi \Sqz_\sigma \Rot_{\phi}^\dagger \nonumber\\
 &= \smallMatrix{\cos \phi & -\sin \phi \\ \sin \phi  & \cos \phi} 
     \smallMatrix{\E{\sigma} & 0 \\ 0  & \E{-\sigma}}
     \smallMatrix{\cos \phi & \sin \phi \\ -\sin \phi  & \cos \phi} ,
\end{align}
which describes squeezing by $\E{-\sigma}$ at angle $\phi$ and
anti-squeezing by $\E{\sigma}$ at $\phi+\pi/2$.  Conventionally,
squeezing magnitudes are expressed in decibels (dB), with \hbox{$\sigma_{dB}
= \sigma \times 20\log_{10}e$}.

In general, all optical losses outside of the filter cavity are
frequency independent or the frequency dependence is so small that it
can be neglected.  Examples of optical losses are residual
transmissions of steering mirrors, scattering, absorption and
imperfections in polarisation optics.  The last of these is likely to
dominate the frequency-independent losses incurred in the passage of
the squeezed field to the readout, therefore these losses are
represented in Figure~\ref{fig:model} as occurring at the optical
isolator.

Since there are no non-linear elements in our system between the squeezer
and the interferometer (i.e. nothing which mixes upper and lower audio
sidebands) we can combine all of the input losses together into a
single frequency-independent ``injection loss'', $\Loss^2_{\rm inj}$,
which represents the total power loss outside of the filter cavity and
before the readout (this work does not consider any losses within the
interferometer itself).

Amalgamating the losses with the action of the squeezer, we arrive at
the two-photon transfer matrix which takes $v_1$ to the filter
cavity\footnote{Our model treats all injection losses as occurring
  before the filter cavity. This approach is valid as the action of
  the filter cavity on coherent vacuum states is null.},
\begin{equation}
\label{eq:inj}
\Tinj = \Attn_{\rm inj}\Sqz(\sigma_\mathrm{sqz}, \phi_\mathrm{sqz}),
\end{equation}
where the attenuation due to $\Loss^2_{\rm inj}$ is described by the
transfer coefficient $\Attn_{\rm inj} = \Attn(\Loss_{\rm inj}) =
\sqrt{1-\Loss^2_{\rm inj}}$.

\subsection{Filter cavity}
\label{sec:filterCavity}

Reflection from a filter cavity is a linear process which can easily
be described in the one-photon, and therefore two-photon, formalisms,
as in equation (A9) of \cite{Eva13a}. However, the approach therein
does not permit one to explore the consequences of filter cavity
imperfections analytically, with resulting loss of physical
insight. Here we revisit this equation and, by making appropriate
approximations, construct a closed-form expression for the action of a
filter cavity in the two-photon formalism.

For a given signal sideband frequency $\Omega$, the complex
reflectivity, $\rfc(\Omega)$, of a filter cavity, using the same
notation as \cite{Eva13a}, is given by
\begin{equation}
\label{eq:rfc}
\rfc(\Omega) = r_{\rm in}  - \frac{t_{\rm in}^2}{r_{\rm in}} \frac{r_{\rm rt} \E{-\I \Phi(\Omega)} }{1- r_{\rm rt} \E{-\I \Phi(\Omega)}} ~,
\end{equation}
where $r_{\rm in}$ is the amplitude reflectivity of the input mirror
and $r_{\rm rt}$ is the cavity's round-trip amplitude reflectivity.
For a cavity of length $L_{\rm fc}$ and resonant frequency $\omfc$,
the round-trip phase $\Phi(\Omega)$ is defined as
 \beq{phi}
\Phi(\Omega) = \left(\Omega - \Delta \omfc\right)\frac{2L_{\rm
    fc}}{c},
\eeq
where $\Delta \omfc  = \omfc - \omega_0$ is the cavity detuning with respect to the
carrier frequency $\omega_0$ and $ c$ is the speed of light.

For a high-finesse cavity near to resonance, we can make the
approximations
\begin{align}
  \label{r_approx}
  \E{-\I \Phi(\Omega)} & \simeq  1 - \I \Phi(\Omega) \\
   {\rm ~ and}\quad r_{\rm rt} &\simeq  r_{\rm in} \simeq \sqrt{1 - t_{\rm in}^2 - \Loss_{\rm rt}^2} \nonumber\\
  & \simeq 1 - (t_{\rm in}^2 + \Loss_{\rm rt}^2) / 2, 
\end{align}
where $\Loss_{\rm rt}^2$ accounts for the power lost during one
round-trip in the cavity (not including input mirror transmission).
 
Under these approximations, and neglecting terms of order 1 or greater
in $\Loss_{\rm rt}^2$, $t_{\rm in}^2$ and $\Phi(\Omega)$,
\eqref{eq:rfc} can be rewritten as \footnote{Note that the $\epsilon$
  defined here is similar to that in equation (94) of \cite{Kim01a}
  and that $\epsilon \to 1$ for an optimally coupled cavity.}
\begin{align}
  \label{eq:rfc_simple}
  \rfc(\Omega) &= 1- \frac{2 - \epsilon}{1 + i
  \xi(\Omega)}
= \frac{\epsilon - 1 +  i \xi(\Omega)}{1 + i \xi(\Omega)},\\
\label{eq:epsilon}
{\rm where}\quad\epsilon &= \frac{2 \Loss_{\rm rt}^2}{t_{\rm in}^2 +
  \Loss_{\rm rt}^2} = \frac{c \; \Loss_{\rm rt}^2}{2 L_{\rm fc} \gfc}
= \frac{f_{\rm FSR}}{\gfc} \; \Loss_{\rm rt}^2,\\
\xi(\Omega) &= \frac{2 \Phi(\Omega)}{t_{\rm in}^2 + \Loss_{\rm
    rt}^2} = \frac{\Omega - \Delta \omfc}{\gfc}
\end{align}
and the cavity half-width-half-maximum-power linewidth is defined as
\begin{equation}
  \label{eq:gfc}
  \gfc=\frac{1-r^2_{\rm rt}}{2}\frac{c}{2L_{\rm fc}}
  =\frac{t^2_{\rm in}+\Loss^{2}_{\rm rt}}{2}\frac{c}{2L_{\rm fc}}.
\end{equation}
As noted by previous authors \cite{Khalili2010}, for a given cavity
half-width $\gfc$, the filter cavity performance is determined
entirely by the loss per unit length $\Loss_{\rm rt}^2 / L_{\rm fc}$.

To investigate the effect the filter cavity has on a squeezed field we
must convert its response, \eqref{eq:rfc_simple}, into the two-photon
picture.  This is done with the one-photon to two-photon conversion
matrix (see \cite{Eva13a} and section \ref{sec:opt}),
\beq{one_to_two} \TwoP = \frac{1}{\sqrt{2}} \smallMatrix{1 & 1 \\ -\I
  &+\I}, \eeq yielding the transfer matrix \beq{TfcA} \Tfc = \TwoP
\cdot \smallMatrix{r_+ & 0 \\ 0 &r_-^*} \cdot \TwoP^{-1}, \eeq where
$r_\pm = \rfc(\pm\Omega)$.

To cast this expression in a more instructive form, we require several
sum and difference quantities based on $ \rfc(\Omega)$. In terms of
$\epsilon$ and $\xi(\Omega)$, the complex phase and magnitude of
$\rfc(\Omega)$ are given by
\begin{align}
   \alpha_{\rm
  fc}(\Omega) &= \tfun{arg}{\rfc(\Omega)}\nonumber\\
  \label{eq:alpha_rho_fc}
&= \tfun{arg}{-1 + \epsilon + \xi^2(\Omega) + \I (2 - \epsilon)\xi(\Omega)} \\
\mathrm{and}\quad\rho_{\rm fc}(\Omega) &= \left| \rfc(\Omega) \right|
 = \sqrt{1 - \frac{(2 - \epsilon)\epsilon}{1 + \xi^2(\Omega)}}.
\end{align}
Whence we define
\begin{alignat}{3}
  \label{eq:alpha_rho}
    \alpha_\pm &=\alpha_{\rm
  fc}(\pm\Omega), & \rho_\pm &=\rho_{\rm fc}(\pm\Omega), \nonumber \\
  \alpha\hspace{.003cm}_{^p_m} &= \frac{\alpha_+ \pm \alpha_-}{2}
&\quad\mathrm{and}\quad\rho\hspace{.003cm}_{^p_m} &= \frac{\rho_+ \pm \rho_-}{2},
\end{alignat}
where the subscripts $p$ and $m$ are used to denote the sum and
difference of the phases and magnitudes.

The transfer matrix of the filter cavity can then be expressed in a
form which clearly shows the effect of intra-cavity loss,
\beq{Tfc}
  \Tfc = \underbrace{\E{\I \alpha_m} \Rot_{\alpha_p} }_{\rm lossless}
  \underbrace{ \left( \rho_p ~ \Eye - \I \rho_m ~ \Rot_{\pi/2} \right)
  }_{\rm lossy},
\eeq
where $\Eye$ is the 2$\times$2 identity matrix.

The first term in this expression, marked ``lossless'', consists of a
rotation operation and an overall phase which are identical to the
rotation and phase provided by a lossless filter cavity \cite{Kim01a}.

The second,``lossy'', term goes to unity for a lossless filter cavity
($\rho_p = 1$ and $\rho_m = 0$). However, in the presence of losses,
this term mixes the quadratures of the squeezed state, corrupting
``squeezing'' with ``anti-squeezing''. We emphasise that this effect
is not \emph{decoherence}, as we have not yet introduced the vacuum
fluctuations which enter as a consequence of the filter cavity losses,
but rather a \emph{coherent dephasing} of the squeezed quadratures
which cannot be undone by rotation of the state. This dephasing is a
direct result of different reflection magnitudes experienced by the
upper and lower audio sidebands (i.e.~$\rho_m \neq 0$). The
ramifications of this effect on the measured noise are presented in
section~\ref{sec:LNT}.

 
Additionally, by combining \eqref{eq:alpha_rho_fc} and
\eqref{eq:alpha_rho}, we are now able to write an explicit expression
for the squeezed quadrature rotation, $\alpha_p$, produced by the filter
cavity,
\begin{equation}
\label{eq:alpha_p}
\alpha_p \simeq \tfun{atan}{\frac{(2 - \epsilon) \gfc \; \Delta\omfc}
 {(1 - \epsilon) \gfc^2 - \Delta\omfc^2 + \Omega^2}},
\end{equation}
which holds for typical filter cavity parameters, $\epsilon \ll 1
\Rightarrow \Loss_{\rm rt}^2 \ll t_{\rm in}^2$. In particular, for a
lossless filter cavity ($\epsilon = 0$), \beq{alpha_p_noloss} \alpha_p
= \tfun{atan}{\frac{2 \gfc \; \Delta\omfc} { \gfc^2 - \Delta\omfc^2 +
    \Omega^2}}, \eeq consistent with the expression for $\alpha_p$
which can be deduced from (88) of \cite{Kim01a} (note that the
referenced equation is missing factor of 2, as reported in
\cite{Har03a}).

\subsection{Mode-matching}
\label{sec:modeMatching}



A quantum filter cavity modifies the phase of the squeezed state which
is coupled into its resonant mode.  In a laboratory context,
free-space optics are used to perform this coupling, maximising the
spatial overlap between the cavity mode and the incident beam. This
process is known as ``mode-matching'' and the result is inevitably
imperfect.  In the case of quantum filter cavities, imperfect
mode-matching results in both a source of loss and in a path by which
the squeezed state can bypass the filter cavity. In this section we
develop a model describing how imperfect filter cavity mode-matching
affects a squeezed state. Furthermore, we also include the effects of
loss arising from mode-mismatch between the squeezed field and the
beam, known as the ``local oscillator'' (LO), used to detect it.

The previously stated filter cavity reflectivity $\rfc$ applies only
to a field perfectly mode-matched to the cavity fundamental mode. In
order to incorporate mode-mismatch, we express the LO and the beam
from the squeezed light source in an orthonormal basis of spatial
modes $U_n$ (e.g. Hermite-Gauss or Laguerre-Gauss modes) such that
\begin{align}
\Usqz &=\sum_{n = 0}^\infty \mmSqz_n  U_n, \quad {\rm with}~~\mmSqz_0  =  \sqrt{1-\sum_{n = 1}^\infty |\mmSqz_n|^2} \\ 
\Ulo &= \sum_{n = 0}^\infty \mmLO_n   U_n, \quad {\rm with}~~\mmLO_0  =  \sqrt{1-\sum_{n = 1}^\infty |\mmLO_n|^2}
\end{align}
where $\mmSqz_n$ and $\mmLO_n$ are complex coefficients.  We further
choose this basis such that $U_0$ is the filter cavity fundamental
mode.
For $\mmSqz_0 = 1$ the beam from the squeezed light source is
perfectly matched to the filter cavity mode.  Similarly, $\mmLO_0 = 1$
indicates that the local oscillator beam has perfectly spatial overlap
with the filter cavity mode.
 
Since the filter cavity is held near the resonance of the fundamental
mode, we assume that all other modes ($U_n$ with $n > 0$) are far from
resonance, with $\xi \gg 1$ and $\rfc \simeq 1$.  Thus, the squeezed
beam after reflection from the filter cavity is given by
\begin{equation}
\Urfc = \rfc(\Omega) \cdot \Usqz = \rfc(\Omega) \; \mmSqz_0 \; U_0
  + \sum_{n = 1}^\infty \mmSqz_n U_n \ .\\
\end{equation}
The fundamental mode's amplitude and phase are modified by the filter
cavity, whereas those of the other modes remain unchanged since these
modes are not resonant and the filter cavity acts like a simple
mirror.

The spatial overlap integral of the reflected field $\Urfc$ and the
local oscillator $\Ulo$ is
\begin{equation}
\label{eq:overlap}
\left< \Ulo | \Urfc\right> = t_{00} ~ \rfc(\Omega) + t_{\rm mm}
\end{equation}
where $t_{00} = \mmSqz_0 \mmLO_0^* {\rm ~ and ~} t_{\rm mm} = \sum_{n = 1}^\infty \mmSqz_n \mmLO_n^*.$
Note that $t_{\rm mm}$ represents the overlap between the
 mismatched part of the beam from the squeezed light source and the mismatched LO.
The squeezed field which follows this path essentially bypasses the filter cavity,
 and thereby \emph{experiences no frequency dependent rotation}.
It may, however, acquire a frequency independent rotation with respect to the
 field which couples into the filter cavity, as can be seen from the two-photon mode-mismatch matrix
\beq{Tmm}
\Tmm = \TwoP \cdot \smallMatrix{t_{\rm mm} & \\ &t_{\rm mm}^*} \cdot \TwoP^{-1} = | t_{\rm mm} | ~ \Rot\big(\tfun{arg}{t_{\rm mm}}\big) \; .
\eeq
The addition of this coupling path results in a frequency
dependent rotation error with respect to the rotation expected from a
perfectly mode matched filter cavity. For modest amounts of
mode-mismatch (less than $10\%$), this error can be corrected by
a small change in the filter cavity detuning.

The magnitude of the mode-mismatch is constrained by $t_{00}$ such
that \beq{MMmax} | t_{\rm mm} | \leq \sqrt{(1 - a_0^2)(1 - b_0^2)}
\leq 1 - t_{00} \eeq while the phase is in general unconstrained. The
$\left< \Ulo | \Urfc\right> $ overlap is maximised when $t_{\rm mm}$
is real and positive and minimised when it is real and negative.

Experimentally, the quantities which one can easily measure are the
squeezed field/filter cavity power mode-coupling, $a_0^2$, and the
squeezed field/local oscillator power mode-coupling, $c_0^2$,
say. From these values one can determine $b_0$, the overlap between
the LO and filter cavity modes, in the following way,
\begin{equation}
\label{eq:modeCoupling}
  b_0 = \left< \Ulo | \Ufc\right>=
  a_0c_0+\sqrt{(1-a_0^2)(1-c_0^2)}\exp(i\phi_{\mathrm{mm}}),
\end{equation}
where $\phi_{\mathrm{mm}}$ captures the ambiguity in the
$t_{\mathrm{mm}}$ phase. The parameters of interest for noise
propagation are then easily determined,
\begin{align}
   t_{00} & = a_0b_0^*,\\
t_{\mathrm{mm}} &= c_0-t_{00}.
\end{align}

Note that the second equality in \eqref{eq:modeCoupling} is not
universally true. The magnitude of the second term (the expression
multiplying the exponential) can be smaller than that given, depending
on the unknown character of the mode-mismatches. However this choice,
an upper bound, allows one to explore the full range of $b_0$ values
necessary to constrain the mode-mismatch-induced noise.


\subsection{Interferometer}
\label{sec:interferometer}
The non-linear action of radiation pressure in an interferometer
affects any vacuum field incident upon it. In our analysis, we include
an idealised lossless interferometer to illustrate this
phenomenon. Operated on resonance, such an interferometer may be
described by the transfer matrix
\begin{equation}
\label{eq:Tifo}
\Tifo = \smallMatrix{1 & 0 \\ -\Kifo & 1},
\end{equation}
as reported in \cite{BnC}. Here, $\Kifo$ characterises the coupling of
amplitude fluctuations introduced at the interferometer's dark port to
phase fluctuations exiting the same port and takes the form
\beq{Kifo}
\Kifo = \left( \frac{\Osql}{\Omega} \right)^2 \frac{\gamma_\ifo^2}{\Omega^2 + \gamma_\ifo^2},
\eeq
where $\gamma_\ifo$ is the interferometer signal-bandwidth and $\Osql$
is a characteristic frequency, dependent on the particular
interferometer configuration, which approximates the frequency
at which the interferometer quantum noise equals the Standard Quantum
Limit (i.e. where radiation pressure noise intersects shot
noise~\cite{Kim01a}).
 
For a conventional interferometer without a signal recycling mirror,
like the power-recycled Michelson interferometer described in
\cite{Kim01a},
\begin{align}
\label{eq:Kifo0}
\gamma_{\ifo_{\,0}}&=\gamma_\arm  \simeq \frac{T_{\rm arm}c}{4L_\mathrm{arm}} \\
\mbox{and}\quad\Omega_{\mathrm{SQL}_{\,0}} & \simeq \frac{8}{c} \sqrt{\frac{P_{\rm arm}\omega_0}{mT_{\rm arm}}},
\end{align}
where $P_{\rm arm}$ is the laser power stored inside the
interferometer arm cavities, $\omega_0$ is the frequency of the
carrier field, $L_{\rm arm}$ is the arm cavity length, $m$ is the mass
of each test mass mirror, $T_{\rm arm}$ is the power transmissivity of
the arm cavity input mirrors and approximations are valid provided arm
cavity finesse is high.

\begin{table}[t!]
\caption{\label{tab:IFOsymbols}Symbols and values for aLIGO
  interferometer parameters.}
\begin{ruledtabular}
\begin{tabular}{lcr}
Parameter & Symbol & Value \\
\hline
Frequency of the carrier field & $\omega_0$ & $2\pi \times$ \SI{282}{THz}\\
Arm cavity length & $L$  & \SI{3995}{m} \\
Signal recycling cavity length & $L_{\rm src}$ & \SI{55}{m}\\
Arm cavity half-width & $\gamma_{\rm arm}$ & $2 \pi \times \SI{42}{\Hz}$\\
Arm cavity input mirror power & \multirow{2}{*}{$T_{\rm arm}$} & \multirow{2}{*}{\SI{1.4}{\%}} \\
transmissivity &&\\
Signal recycling mirror power & \multirow{2}{*}{$t^2_{\rm sr}$} & \multirow{2}{*}{$ \SI{35}{\%}$}\\
transmissivity &&\\
Intra-cavity power & $P_{\rm arm}$ & \SI{800}{kW} \\
Mass of each of the test mass mirror & $m$ & \SI{40}{kg}
\end{tabular}
\end{ruledtabular}
\end{table}

For a dual-recycled interferometer, operating with a tuned
signal-recycling cavity of length $L_{\rm src}$, it can be shown that,
for $\Omega \ll c / L_{\rm src}$,
\begin{align}
\label{OmegaSQL}
\gamma_\ifo &= \frac{1 + r_{\rm sr}}{1 - r_{\rm sr}} \gamma_{\ifo_{\,0}} \\
\mbox{and}\quad\Osql &= \frac{t_{\rm sr}}{1 + r_{\rm sr}} \Omega_{\mathrm{SQL}_{\,0}},
\end{align}
where $t_{\rm sr}$ and $r_{\rm sr}$ are the amplitude transmissivity
and reflectivity of the signal recycling mirror.  Given the Advanced
LIGO parameters reported in Table~\ref{tab:IFOsymbols},
 \begin{alignat}{3}
\label{eq:OMEGASQLnum}
 \gamma_\ifo &\simeq \:\: 9~\gamma_{\ifo_{\,0}}  &&\simeq  2 \pi \times \SI{390}{\Hz}\\
 \mbox{and}\quad\Osql &\simeq~\frac{\Omega_{\mathrm{SQL}_{\,0}}}{3} &&  \simeq 2 \pi \times \SI{70}{\Hz},
 \end{alignat}
 confirming that the effect of signal recycling in Advanced LIGO is to
 increase the interferometer's bandwidth whilst reducing the frequency
 at which its quantum noise reaches the SQL. For such an
 interferometer, in which $\gamma_\ifo \gg \Osql$, $\Kifo$ may be
 approximated by $(\Osql / \Omega)^2$ in the region of interest (where
 $\Kifo$ is order unity or larger).

While \eqref{eq:Tifo} is very simple, greater appreciation of the
action of the interferometer can be gained by noting that $\Tifo$ can
be recast in terms of the previously defined squeeze and rotation
operators as
\begin{gather}
  \elabel{Tifo2}
  \Tifo = \Sqz(\sigma_\ifo,\phi_{\rm ifo}) \Rot(\theta_\ifo)
\end{gather}
with
\begin{align}
  \sigma_\ifo &= -\arcsinh (\Kifo / 2),\nonumber\\
  \phi_\ifo &= \tfrac{1}{2}\arccot (\Kifo / 2), \nonumber\\
\theta_\ifo &= -\arctan (\Kifo / 2). \nonumber 
\end{align}
The role of the filter cavity is to rotate the input squeezed
quadrature as a function of frequency such that it is always aligned
with the signal quadrature at the output of the interferometer, even
in the presence of rotation by $\theta_\mathrm{ifo}$ and the effective
rotation caused by squeezing at angle $\phi_\mathrm{ifo}$. The
required filter cavity rotation is given by
\begin{equation}
  \label{eq:fcRot}
  \theta_\mathrm{fc} = \arctan(\Kifo).
\end{equation}

\subsection{Linear noise transfer}
\label{sec:LNT}
We now combine the intermediate results of previous sections to
compute the quantum noise observed in the interferometer
readout. Three vacuum fields make contributions to this noise: $v_1$
which passes through the squeezer, $v_2$ which enters before the
interferometer but does not pass through the squeezer and $v_3$ which
enters after the interferometer. We formulate transfer matrices for
each of these fields in turn before providing, in \eqref{eq:noise}, a
final expression for the measured noise.

Converting the result of \eqref{eq:overlap} into a two-photon transfer
matrix and including losses in the injection and readout paths, via
$\Tinj$ and $\tau_\mathrm{ro}$ respectively (see \eqref{eq:inj} and
\eqref{eq:tau3}), we arrive at the full expression describing the
transfer of vacuum field $v_1$ through the squeezer, filter cavity and
interferometer to the detection point,
\begin{equation}
  \label{eq:Tsqz}
  \Tsqz = \Attn_{\rm ro} \Tifo \left( t_{00} ~ \Tfc + \Tmm \right) \Tinj.
\end{equation}

We now consider the vacuum field $v_2$, which accounts for all
fluctuations coupled into the beam due to injection losses, losses
inside the filter cavity itself and imperfect mode-matching. The
audio-sideband transmission coefficient from the squeezer to the
interferometer is
 \beq{tau} \tau_2(\Omega) = (t_{00} \; \rfc(\Omega) +
t_{\rm mm}) \Attn_{\rm inj} .  \eeq
In the two-photon picture, the average of the upper and lower sideband
losses gives the source term for the $v_2$ vacuum fluctuations, so that
\begin{align}
\label{eq:T2A}
  \TnN{2} &= \Attn_{\rm ro} \Tifo \Loss_2\\
  \label{eq:T2B}
  \mathrm{where}\quad\Loss_2 &= \sqrt{1- (| \tau_2(+\Omega) |^2 + | \tau_2(-\Omega) |^2) / 2}.
\end{align}

Finally, frequency independent losses $\Lambda^2_{\rm ro}$ between the
interferometer and the readout introduce a second source of
attenuation of the squeezed state and accompanying vacuum fluctuations
$v_3$, a process described by the following transfer matrix and
transmission coefficient
\begin{equation}
  \label{eq:tau3}
\TnN{3} = \Lambda_{\rm ro}, \quad  \Attn_{\rm ro} = \sqrt{1-\Lambda^2_{\rm ro}}.
\end{equation}
These losses cannot be added to the injection losses mentioned above
since they are separated by the non-linear effects of the
interferometer. Explicitly, losses before and after the interferometer
are not equivalent.

The single-sided power spectrum of the quantum noise at the
interferometer readout is then given by
\begin{gather}
\label{eq:noise}
N(\zeta) = \underbrace{\left| \Fhd \cdot \Tsqz \cdot v_1 \right|^2 }_{N_{1}(\zeta)}+
\underbrace{\left| \Fhd \cdot \TnN{2}  \cdot v_2 \right|^2} _{N_{2}(\zeta)}+
 \underbrace{\left| \Fhd \cdot \TnN{3}  \cdot v_3 \right|^2}_{N_{3}(\zeta)}
\end{gather}
where the local oscillator field $\Fhd = A_{\rm LO} \smallMatrix{
  \sin\zeta & \cos\zeta }$, with amplitude $A_{\rm LO}$,
determines the readout quadrature. All mathematical operations are as
defined in \cite{Eva13a} and $\zeta$ is defined such that $N(\zeta =
0)$ is the noise in the quadrature containing the interferometer
signal.

We now investigate \eqref{eq:noise} more closely, providing analytical
expressions for the contribution of each term. To improve readability,
we normalise all noise powers with respect to shot noise (see
appendix~ \ref{sec:calc}). This action is denoted through the use of
an additional circumflex, i.e.~$\widehat{N}$ rather than $N$.

\subsubsection{Noise due to vacuum fluctuations passing through the
  squeezer, $\widehat{N}_1$}

As the only term with dependence on filter cavity performance,
examination of $\widehat{N}_{1}(\zeta)$ allows one to determine the optimal
filter cavity parameters.

A comprehensive expression for $\widehat{N}_{1}(\zeta = 0)$ may be developed
starting from \eqref{eq:Tsqz}. However, for clarity, and to assist in
gaining physical understanding, we restrict our discussion to an
optimally matched filter cavity, and neglect injection and
  readout losses, to obtain a simple description in terms of the
optomechanical coupling constant $\Kifo$, the cavity rotation angle
$\alpha_p$ and reflectivities $\rho_p$ and $\rho_m$. In this case,
\begin{align}
  \widehat{N}_{1}(\zeta = 0)  = &\left(\rho_p^2e^{-2 \sigma}  +
    \rho_m^2 e^{2 \sigma}\right)
  \left(\cos{\alpha_p}+\Kifo\sin{\alpha_p}\right)^2  \nonumber\\
  \label{eq:noise1}
   +  &\left(\rho_p^2e^{2 \sigma}+ \rho_m^2 e^{-2 \sigma}\right)\left(\Kifo\cos{\alpha_p}-\sin{\alpha_p}\right)^2.
\end{align}

Equation \eqref{eq:noise1} elucidates both the effect of a filter cavity
and the role of filter cavity losses.  We first remark that in the
absence of both squeezed light ($\sigma=0$) and a filter cavity
(equivalent to $\rho_p = 1$, $\rho_m = 0$) the interferometer output
noise is simply
\begin{equation}
  \widehat{N}_{1} = 1 + \Kifo^2.
\end{equation}

With the addition of frequency-independent squeezed light
($\sigma\neq0$, $\alpha_p = 0$), the total output noise becomes
\begin{equation}
\widehat{N}_{1} = e^{-2 \sigma}+ e^{2 \sigma}\Kifo^2.
\end{equation}
In the frequency region in which $\Kifo < 1$ the noise is reduced by
the presence of squeezed light but for $\Kifo > 1$ the noise is
degraded by the ``anti-squeezing'' component $e^{2 \sigma}$. Had we
chosen $\alpha_p = \pi/2$ these roles would have been reversed.
 
The presence of a filter cavity ($\alpha_p=\alpha_p(\Omega) \neq 0$)
allows one to minimise the impact of ``anti-squeezing'' on the
measured noise.  For a lossless filter cavity ($\rho_m = 0$, $\rho_p =
1$) the ``anti-squeezing'' can be completely nulled by selecting
filter cavity parameters such that $\alpha_p =
\arctan(\Kifo)$, giving the minimal quantum noise
\begin{equation}
\widehat{N}_{1} = e^{-2 \sigma}\left(1
  +\Kifo^2\right).
\end{equation}

With the addition of filter cavity losses ($\rho_m \neq 0$) the total
noise becomes
\begin{equation}
\widehat{N}_{1}= \left(\rho^2_p e^{-2
    \sigma}+\rho^2_m e^{2 \sigma}\right)\left(1
  +\Kifo^2\right)
\end{equation}
and there is no value of $\alpha_p$ for which the influence of
``anti-squeezing'' can be completely nulled (due to the coherent
dephasing effect discussed above in \ref{sec:filterCavity}). It is
important to highlight that precluding ``optimal'' rotation is not the
only downside of a lossy filter cavity. Intra-cavity losses also
introduce additional vacuum fluctuations, $v_2$, which do not pass
through the squeezer, leading to increased noise in the interferometer
readout via the $T_2$ transfer matrix. Considering an optimally
mode-matched filter cavity, this effect is most noticeable in
\eqref{eq:T2B}, which becomes simply $\Lambda_2 = \sqrt{1-(\rho^2_p+\rho^2_m)}
\neq 0$ (see also section \ref{sec:N2} below).

For an interferometer in which $\gamma_\ifo \lesssim \Osql$, like a
power-recycled Michelson interferometer (or a detuned signal-recycled
Michelson interferometer), a single filter cavity is not capable of
realising the desired rotation of the squeezed quadrature, as
extensively described in section V and appendix C of
\cite{Kim01a}. Conversely, for a broadband interferometer like
Advanced LIGO, in which $\gamma_\ifo > 5~\Osql$ and the approximation
$\Kifo \simeq (\Osql / \Omega)^2$ holds, it can be shown, from
\eqref{eq:alpha_p} and \eqref{eq:fcRot}, that the output noise is
minimised by a single filter cavity with the following parameters
\begin{align}
\label{eq:fc_rotA}
 \Delta \omfc & =  \sqrt{1 - \epsilon}~\gfc\\
 \label{eq:fc_rotB}
 \mbox{and}\quad\gfc & = \sqrt{\frac{2}{(2-\epsilon)\sqrt{1-\epsilon}}}~\frac{\Osql}{\sqrt{2}},
\end{align}
from which the requirements for a lossless filter cavity ($\epsilon =
0$) can be derived,
\begin{align}
  \label{eq:fc_rot_noLossA}
  \Delta \omfc & = \gfc \\
  \label{eq:fc_rot_noLossB}
  \mbox{and}\quad \gfc  & = \frac{\Osql}{\sqrt{2}}.
\end{align}

In practice, for fixed cavity length and losses, the value of
$t_\mathrm{in}$ is tuned to obtain the required filter cavity
bandwidth. However, changing $t_\mathrm{in}$ affects both $\epsilon$
and $\gfc$, making \eqref{eq:fc_rotB} inconvenient to
solve. Nevertheless, equating the right-hand side of
\eqref{eq:fc_rotB} with the expression for $\gfc$ derived from
\eqref{eq:epsilon}, one obtains a version of $\epsilon$ which is
independent of $t_\mathrm{in}$,
\begin{equation}
  \epsilon=\frac{4}{2+\sqrt{2+2\sqrt{1+\left(\frac{2 \Osql}{f_{\rm
              FSR}\Loss_{\rm rt}^2}\right)^4}}},
\end{equation}
and can be used to find $\Delta \omfc$ and $\gfc$. Then, from
\eqref{eq:gfc},
\begin{equation}
  t_\mathrm{in}^2=\frac{2 \gfc}{f_{\rm FSR}}-\Loss_{\rm rt}^2.
\end{equation}
We note that as filter cavity losses increase, the ideal filter
cavity bandwidth also increases, whilst the optimal cavity detuning is
reduced. As a consequence, the desired value of $t_\mathrm{in}$ is
approximately constant for $\epsilon\lesssim0.3$.

\subsubsection{Noise due to vacuum fluctuations which do not pass
  through the squeezer, $\widehat{N}_2$}
\label{sec:N2}

Let us now consider $\widehat{N}_{2}(\zeta = 0)$, the term
describing noise due to loss-induced vacuum fluctuations which do not
pass through the squeezer. Assuming perfect mode-matching, $\Lambda_2$
from \eqref{eq:T2B} can be written as
\begin{align}\label{eq:noise2}
\Lambda_2 & = \sqrt{1-\tau^2_{\rm inj}\left(\rho^2_p + \rho^2_m \right)}.
\end{align}
Thus, using \eqref{eq:T2A}, we obtain
\begin{align}
  \widehat{N}_{2}(\zeta = 0) & = \Attn^2_{\rm ro}\left(1+\Kifo^2\right)\Lambda^2_2
  \nonumber \\
  \label{eq:noise2}
  & = \Attn^2_{\rm ro}\left(1+\Kifo^2\right)\left(1-\tau^2_{\rm inj}\left(\rho^2_p +
      \rho^2_m \right)\right).
\end{align}

\subsubsection{Noise due to vacuum fluctuations in the readout, $\widehat{N}_3$}

The noise due to vacuum fluctuations entering at the interferometer
readout follows trivially from \eqref{eq:tau3},
 \begin{align}\label{eq:noise2}
   \widehat{N}_{3}(\zeta = 0) & = \Lambda^2_{\rm ro} =
   1-\Attn^2_{\rm ro}.
 \end{align}

\subsection{Phase noise}
\label{sec:phaseNoise}

In addition to optical losses and mode-mismatch, a further cause of
squeezing degradation is phase noise, also referred to as ``squeezed
quadrature fluctuations''~\cite{Dwy13}. In this section we develop a
means of quantifying the impact of this important degradation
mechanism.

Assuming some parameter $X$ in $\TnN{1}$ or $\TnN{2}$ has small,
Gaussian-distributed fluctuations with variance $\delta X^2$, the
average readout noise is given by
\begin{align}
\widehat{N}_{\rm avg}(\zeta) &\simeq \widehat{N}(\zeta, X) +  \partderivS{\widehat{N}(\zeta, X)}{X} ~ \frac{\delta X^2}{2} \\
&\simeq \frac{\widehat{N}(\zeta, X + \delta X) +  \widehat{N}(\zeta, X - \delta X)}{2}.
\end{align}
Extending this approach to multiple incoherent noise parameters $X_n$
yields
\begin{align}
  \widehat{N}_{\rm avg} &\simeq \widehat{N} + 
  \sum_n\partderivS{\widehat{N}(X_n)}{X_n} ~ \frac{\delta X_n^2}{2} \\
  &\simeq \widehat{N} + \sum_n \left( \frac{\widehat{N}(X_n + \delta X_n) +  \widehat{N}(X_n - \delta X_n)}{2} - \widehat{N} \right),
  \elabel{phaseNoise}
\end{align}
where the parameters not explicitly listed as arguments to $\widehat{N}$,
including $\zeta$, are assumed to take on their mean values.

While (\eref{phaseNoise}) is sufficient to evaluate $\widehat{N}_\mathrm{avg}$
for any collection of phase noise sources, we choose to follow the
same approach adopted in the treatment of optical losses, considering
two classes of squeezed quadrature fluctuations: extra-cavity
fluctuations which are frequency independent and intra-cavity
fluctuations which are frequency dependent.

Examples of frequency-independent phase noise sources include length
fluctuations in the squeezed field injection path and instabilities in
the relative phase of the local oscillator or the radio-frequency
sidebands which co-propagate with the squeezed field. Such
frequency-independent noise may be represented by variations,
$\delta\zeta$, in the homodyne readout angle $\zeta$.

Frequency-dependent phase noise is caused by variability in the filter
cavity detuning $\Delta\omfc$ (see (\eref{phi})). This detuning noise
results from filter cavity length noise $\delta L_{\rm fc}$, driven by
seismic excitation of the cavity mirrors or sensor noise associated
with the filter cavity length control loop, according to
\beq{lengthToDetuning} \delta\Delta\omfc = \frac{\omega_0}{L_{\rm fc}}
\delta L_{\rm fc}. \eeq

Detuning noise gives rise to frequency-dependent phase noise through
the properties of the filter cavity resonance. For example, the
dependence of $\Tfc$ on $\Delta\omfc$ is weak for $\Omega \gg
\Delta\omfc$, i.e.~for frequencies far from resonance, and stronger
for $\Omega \simeq \Delta\omfc$, i.e.~for frequencies close to
resonance.

General analytic expressions for $\widehat{N}_\mathrm{avg}$ as a function of
$\delta\zeta$ and $\delta L_{\rm fc}$ are neither concise nor
especially edifying. Therefore, in the following section, we apply
\eqref{eqn:phaseNoise} numerically to illustrate the impact of phase
noise in a typical advanced gravitational-wave detector.

\section{A 16 m filter cavity for Advanced LIGO}
\label{sec:param}

We now apply the analytical model expounded above to the particular
case of a 16 m filter cavity. Such a system has recently been
considered for application to Advanced LIGO~\cite{Eva13a} and
therefore we use the specifications of this interferometer in our
study (see Table~\ref{tab:IFOsymbols}).

%

The remaining parameters, show in Table~\ref{tab:param}, represent
what we believe is technically feasible using currently available
technology.  For example, the filter cavity length noise estimate
$\delta L_{\rm fc}$ assumes that the cavity mirrors will be held in
single-stage suspension systems located on seismically isolated
HAM-ISI tables~\cite{Eva12a} and that the filter cavity length control
loop will have \unit[150]{Hz} unity gain frequency, and whilst a 2\%
mode-mismatch between the squeezed field and the filter cavity is
extremely small, newly developed actuators \cite{Kas13a,Liu13a} allow
us to be optimistic. We chose to inject \SI{9.1}{dB} of squeezing into
our system as this value results in \SI{6}{dB} of high-frequency
squeezing at the interferometer readout (a goal for second-generation
interferometers \cite{Bar13a}) and, conservatively, to consider a
filter cavity with 16 ppm round-trip loss, even if recent
investigations have shown that lower losses are achievable
\cite{Isogai13}.

\begin{table}
\caption{\label{tab:param}Parameters used in in the application of our
  model to Advanced LIGO}
\begin{ruledtabular}
\begin{tabular}{p{5.25cm}cr}
Parameter & Symbol & Value \\
\hline
Filter cavity length & $L_{\rm fc}$  & \SI{16}{m} \\
Filter cavity half-bandwidth & $\gfc$ & $2 \pi \times \SI{61.4}{\Hz}$\\
Filter cavity detuning & $\Delta \omfc$ & $2 \pi \times \SI{48}{\Hz}$\\
Filter cavity input & \multirow{2}{*}{$t_{\rm in}^2$} &
\multirow{2}{*}{\SI{66.3}{ppm}}\\
mirror transmissivity &&\\
Filter cavity losses & $\Loss_{\rm rt}^2$ &  \SI{16}{ppm}\\
Injection losses & $\Loss_{\rm inj}^2$ & $5\%$ \\
Readout losses & $\Loss_{\rm ro}^2$ & $5\%$ \\
Mode-mismatch losses & \multirow{2}{*}{$\Loss_{\rm mmFC}^2$} & \multirow{2}{*}{$2\%$} \\
(squeezer-filter cavity) &&\\
Mode-mismatch losses & \multirow{2}{*}{$\Loss_{\rm mmLO}^2$} & \multirow{2}{*}{$5\%$} \\
(squeezer-local oscillator) &&\\
Frequency independent phase noise (RMS) &  $\delta\zeta$  & \SI{30}{mrad} \\
Filter cavity length noise (RMS)& $\delta L_{\rm fc}$& $\SI{0.3}{pm}$\\
Injected squeezing & $\sigma_\mathrm{dB} $& $\SI{9.1}{dB} $
\end{tabular}
\end{ruledtabular}
\end{table}

\begin{figure*}[t!]
    \includegraphics[width=0.85\textwidth]{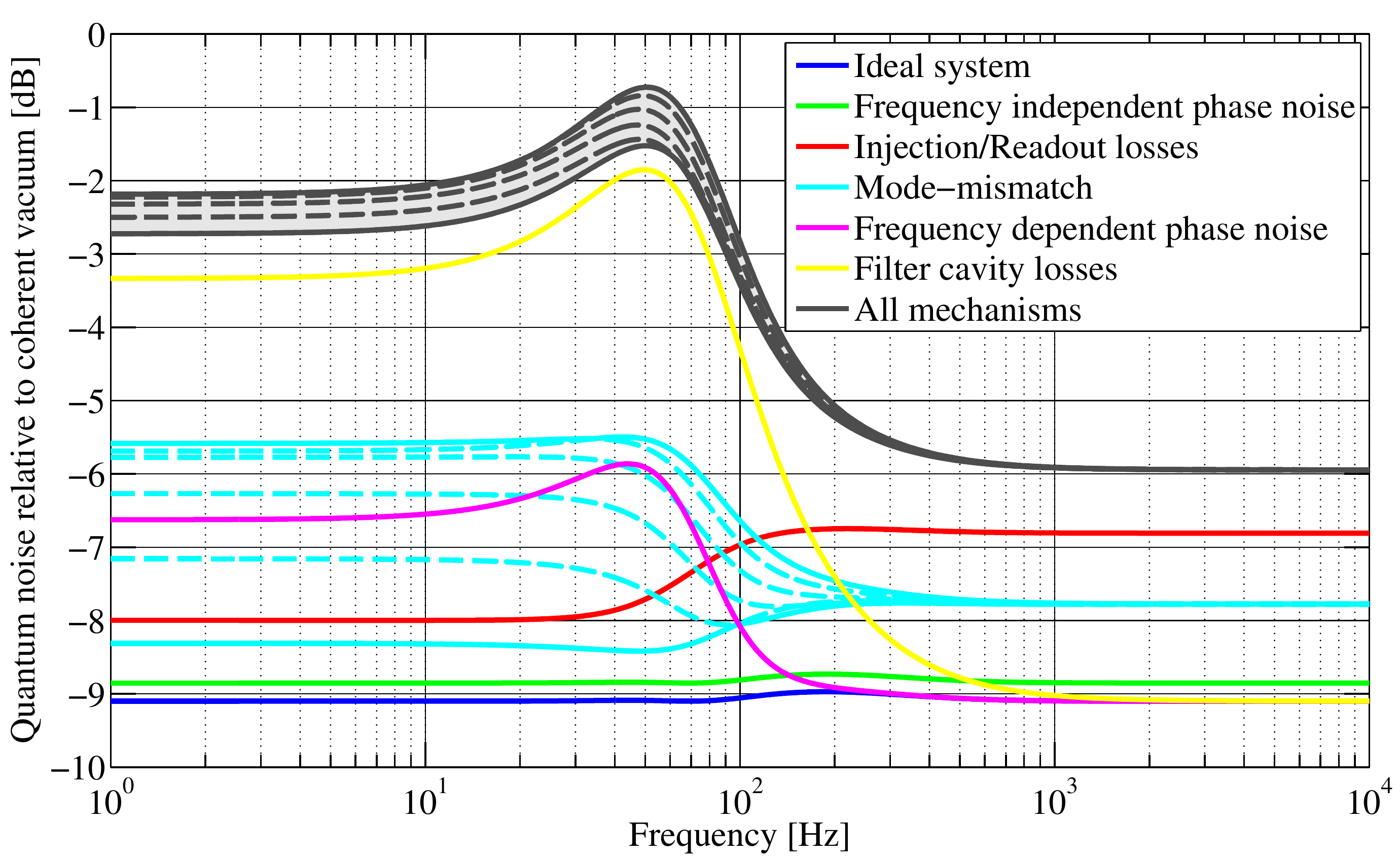}
    \caption{Power spectral density of quantum noise in the signal
      quadrature relative to coherent vacuum. Traces show how the
      noise reduction of a \SI{-9.1}{dB} minimum uncertainty squeezed
      state is impaired by each of the various decoherence and
      degradation mechanisms discussed herein.  The effects of
      coherent dephasing are included in the `Filter cavity losses'
      trace. The family of `Mode-mismatch' curves encapsulates the
      unknown phase of $t_\mathrm{mm}$, with solid curves defining
      upper and lower bounds for the induced noise (see section
      \ref{sec:modeMatching}, specifically
      \eqref{eq:modeCoupling}). The trace labelled `All mechanisms'
      illustrates the total impact when the contributions of all
      decoherence and degradation effects are considered
      simultaneously.}
\label{fig:result}
\end{figure*}

The results of our investigation are shown in
Figure~\ref{fig:result}. One observes that intra-cavity losses are the
dominant source of decoherence below $\sim$\SI{300}{Hz}. 
However, we
note that, with small changes in parameter choice, the impact of the
other coupling mechanisms could also become important. For instance,
filter cavity length fluctuations approaching \SI{1}{pm} RMS would greatly
compromise low frequency performance. 

At higher frequencies, injection, readout and mode-mismatch losses are
the most influential effects. With total losses of $\sim$15\%,
measuring \SI{6}{dB} of squeezing demands that more than 9 dB be
present at the injection point. 

Even under the idealised condition of negligible filter cavity losses
($\Loss_{\rm rt}^2/L_\mathrm{fc}\ll$1ppm/m), achieving a broadband
improvement greater than \SI{6}{dB} places extremely stringent
requirements on the mode-matching throughout the system and on the
filter cavity length noise.

\section{Conclusions}
Quantum filter cavities were proposed several years ago as means of
maximising the benefit available from squeezing in advanced
interferometric gravitational-wave detectors \cite{Kim01a}. However,
the technical noise sources which practically limit filter cavity
performance have, until now, been neglected. In this paper we have
presented an analytical model capable of quantifying the impact of
several such noise sources, including optical loss, mode-mismatch and
frequency dependent phase noise. We find that real-world decoherence
and degradation can be significant and therefore must be taken into
account when evaluating the overall performance of a filter cavity.
Applying our model to the specific case of Advanced LIGO
\cite{Eva13a}, we conclude that a 16 m filter cavity, built with
currently available technology, offers considerable performance gains
and remains a viable and worthwhile near-term upgrade to the
generation of gravitational-wave detectors presently under
construction.

\appendix
\section{Formalism}
\label{app:Formalism}

In this appendix we place the calculations presented above in the
context of the one-photon and two-photon formalisms extensively
discussed in literature (see e.g.~\cite{Cor05a,BnC}). We commence by
connecting the one-photon expression for the time-varying part of the
electromagnetic field to power fluctuations on a photo-detector. We
then transform the derived expression into the two-photon basis to
explicitly show how vacuum fluctuations generate measurable
noise. This calculation is subsequently generalised to the case of
multiple vacuum fields arriving at a photo-detector after having
propagated through an optical system, revealing the origin of
\eqref{eq:noise}. Finally, we discuss how quantum noise may be
calculated for systems best described in the one-photon picture, in
the process deriving the one-photon to two-photon conversion matrix
\eqref{eqn:one_to_two}.

\subsection{One-photon and two-photon in context}

The one-photon and two-photon formalisms provide two alternative ways
of expressing fields. In the one-photon formalism, as described by
(2.6) of~\cite{BnC}, the time varying part of the electromagnetic
field $E(t)$ is written in terms of its audio-sideband components
around the carrier frequency $\omega_0$,
\begin{widetext}
\begin{align}
  E(t) &= \sqrt{\frac{2\pi\hbar \omega_0}{\sA c}}e^{-i\omega_0 t}
   \int_0^{+\infty} \left[a_+(\Omega)e^{-i\Omega t} +
    a_-(\Omega)e^{i\Omega t}\right]\frac{d\Omega}{2\pi} + \mbox{h.c.}\nonumber \\
 &= \sqrt{\frac{2 \pi \hbar \omega_0}{\sA c}} \cdot
  2 \real{e^{-i \omega_0 t} \int_{0}^{+\infty}\!\!\! \left[a_+(\Omega) e^{-i \Omega t} + 
 a_-(\Omega) e^{i \Omega t}\right] \; \frac{d\Omega}{2 \pi} },
\end{align}
\end{widetext}
where $\sA$ is the ``effective area'', ``h.c.'' means Hermitian
conjugate and $a_\pm(\Omega)$ are the normalised amplitudes of the
upper and lower sidebands at frequencies $\omega_0 \pm \Omega $ in
dimensions of $(\mbox{number of photons}/\mathrm{Hz})^{1/2}$ (see
\cite{Cav85a} for greater detail).



By introducing  $\sE$ defined as
\begin{equation}
  \sE  = \sqrt{\frac{2}{\sA c \epsilon_0}}, 
\end{equation}
and noting that \cite{BnC} uses $\epsilon_0 = 1 / 4 \pi $, $E(t)$ can
be rewritten as
\begin{align}
E(t) = &\sE \sqrt{\hbar \omega_0}\nonumber\\
&\times  \real{e^{-i \omega_0 t} \int_{0}^{+\infty}\!\!\! \left[a_+(\Omega) e^{-i \Omega t} + 
 a_-(\Omega) e^{i \Omega t}\right] \; \frac{d\Omega}{2 \pi} }\nonumber \\
 =   &\real{\sE  \; \dA(t) \; e^{-i \omega_0 t}} 
\end{align}
where we have introduced the time-dependent amplitude
\beq{dAonePhoton} \dA(t) = \sqrt{\hbar \omega_0}
\int_{0}^{+\infty}\!\!\! \left[a_+(\Omega) e^{-i \Omega t} +
  a_-(\Omega) e^{i \Omega t}\right] \; \frac{d\Omega}{2 \pi}. \eeq

In our application these fluctuations arrive to the photo-detector
together with a strong, constant \emph{local oscillator} field $A_0$
such that \beq{EwithA0} E(t) = \real{\sE \; (A_0 + \dA(t)) \; e^{-i
    \omega_0 t}}.  \eeq The power $P(t)$ transported by the beam can
then be written as
\begin{align}
P(t) &=  \sA \overline{I(t)}=\sA c \epsilon_0 \, \overline{E(t)^2}
 = \frac{\sA c \epsilon_0}{2} \abs{\sE \left( A_0 + \dA(t) \right) }^2 \nonumber\\
  &= \abs{A_0}^2 + 2 \real{A_0^* \; \dA(t)} + \abs{\dA(t)}^2,
\end{align}
where $I(t)$ denotes intensity and the overbar indicates the average
over one or more cycles of the electromagnetic wave. Note that the
effective area $\sA$ has cancelled and does not have a meaningful
effect on the measurable power. Since $\dA(t) \ll A_0$, we can
approximate the power fluctuation $\dP(t)$ as \beq{dP} \dP(t) \equiv
P(t) - \abs{A_0}^2 \simeq 2 \real{A_0^* \; \dA(t)}.  \eeq

Switching to the frequency domain, we take the Fourier transform of
$\dP(t)$ to find
\begin{align}
\dPf(\Omega) &=  \int_{-\infty}^{+\infty}\!\!\! 2 \real{A_0^* \; \dA(t)} e^{i \Omega t} dt \nonumber\\
 &=  \int_{-\infty}^{+\infty} \left[ A_0^* \; \dA(t) + A_0 \; \dA^*(t) \right] e^{i \Omega t} dt \nonumber\\
 &=  A_0^* \; \dAf(\Omega) + A_0 \; \dAf^*(-\Omega) \nonumber\\
 &= \sqrt{\hbar \omega_0} \left[ A_0^* \; a_+(\Omega) +
  A_0 \; a_-^*(\Omega) \right]\label{eq:dPlast}
\end{align}
 where, in the final step, we have used (\eref{dAonePhoton}).

The two-photon formalism defines quadrature fields as linear combinations of the one-photon fields~\cite{BnC}
\beq{A2}
 a_1 = \frac{(a_+ + a_-^*)}{\sqrt{2}}\quad \mbox{and} \quad a_2 = \frac{(a_+ - a_-^*)}{\sqrt{2}i}
\eeq
 such that
\beq{A2inv}
 a_+ = \frac{(a_1 + i a_2)}{\sqrt{2}} \quad \mbox{and} \quad a_-^* = \frac{(a_1 - i a_2)}{\sqrt{2}}.
\eeq
By substituting (\eref{A2inv}) into \eqref{eq:dPlast}, we obtain the frequency-domain expression for $\dP$ in the two-photon formalism
\begin{align}
\dPf(\Omega) &= \sqrt{\hbar \omega_0/2} \left[ (A_0^* + A_0) a_1(\Omega) + 
  i (A_0^* - A_0) a_2(\Omega) \right] \nonumber\\
 &= \sqrt{2 \hbar \omega_0} \left[ \real{A_0} a_1(\Omega) + \imag{A_0} a_2(\Omega) \right]. 
\end{align}

Expressing the local oscillator's amplitude and phase explicitly, $A_0
= A_{\rm LO} e^{i \phi}$, $\dPf(\Omega)$ becomes \beq{dPf_two}
\dPf(\Omega) = \sqrt{2 \hbar \omega_0}~A_{\rm LO} \left[ a_1(\Omega)
  \cos \phi + a_2(\Omega) \sin \phi \right]. \eeq


\subsection{Calculation of quantum noise}
\label{sec:calc}
 
Equation (\eref{dPf_two}) provides a simple method of calculating the
power fluctuations on a photo-detector given any time-varying
electromagnetic field beating against a local oscillator.

As a specific and relevant example, quantum noise (due to the
zero-point energy of the electromagnetic field) drives vacuum
fluctuations, $a_1(\Omega)$ and $a_2(\Omega)$, which are incoherent
and of unit amplitude at all frequencies. The resulting noise power
generated being
\begin{align}
N = |\dPf|^2 &= 2 \hbar \omega_0 A^2_\mathrm{LO} (\abs{a_1 \cos\phi}^2 + \abs{a_2 \sin\phi}^2)\\
 &= 2 \hbar \omega_0 A^2_\mathrm{LO},
\end{align}
where $a_1$ and $a_2$ have initially been listed explicitly to
highlight the incoherent nature of the noise associated with each of
the two quadratures. Note that this expression is consistent with the
familiar equation $\sqrt{2 P_\mathrm{avg} h \nu}$ for the amplitude
spectral density of shot noise, since the average power level
$P_\mathrm{avg}$ is equal to $A^2_\mathrm{LO}$.

The tools of linear algebra can now be exploited to simplify these
expressions, allowing one to rewrite the noise as
\begin{equation}
N = \abs{A_\mathrm{LO} \smallMatrix{\cos\phi & \sin\phi} \cdot
  \sqrt{2 \hbar \omega_0} \smallMatrix{1 & 0 \\ 0 & 1}}^2 =
  \abs{\Fhd \cdot v_{in}}^2,
\end{equation}
where the local oscillator is as defined in section~\ref{sec:LNT}
(given the LO phase convention $\zeta = \pi / 2 - \phi$),
\begin{equation}
\Fhd = A_{\rm LO} \smallMatrix{\sin\zeta & \cos\zeta }
  = A_{\rm LO} \smallMatrix{\cos\phi & \sin\phi },
\end{equation}
and $v_{in}$,
simply proportional to the $2\times2$ identity matrix, embodies the
two independent vacuum noise sources
\begin{equation}
v_{in} = \sqrt{2\hbar\omega} \; \Eye.
\end{equation}

In general, to calculate the quantum noise in an optical system, the
vacuum field $v_{\mathrm{in}}$ entering an open port is propagated to
the readout photodetector through the transfer matrix $\mathbf{T}$ of
the system,
\begin{equation}
v_{\mathrm{out}}=\mathbf{T} \cdot v_{\mathrm{in}},
\end{equation}
as described in \cite{Eva13a}. The vacuum fluctuations
$v_{\mathrm{out}}$ then beat against the local oscillator field
present on the photodetector to give the power spectrum of quantum
noise
\begin{equation}
N = \abs{\Fhd \cdot v_{out}}^2 = \abs{\Fhd \cdot \TnN{} \cdot v_{in}}^2.
\end{equation}

If multiple paths lead to the same photodetector, the total noise may
be calculated as the sum of the contributions due to each vacuum
source,
\begin{equation}
N = \sum_n \abs{\Fhd \cdot \Tn \cdot v_n}^2 = 2 \hbar \omega_0 \sum_n \abs{\Fhd \cdot \Tn}^2.
\end{equation}
Finally, dividing by the shot noise level gives the normalised noise
power used throughout this paper
\begin{equation}
  \widehat{N} = \frac{N}{2 \hbar \omega_0 A_\mathrm{LO}^2}.
\end{equation}

\subsection{One-photon transfer}
\label{sec:opt}
Some optical systems, like filter cavities, are better described by
the one-photon formalism, as this makes their transfer matrices
diagonal. As in the two-photon formalism, the quantum noise $N$ is the
result of the incoherent sum of the noise generated by two vacuum
fields. Although, in this case, the fields of concern are $a_+$ and
$a_-$ (rather than $a_1$ and $a_2$). Beginning from \eqref{eq:dPlast},
the resulting noise is
\begin{equation}
  N = |\dPf|^2 =\hbar \omega_0 (\abs{A_0^* \, a_+}^2 + \abs{A_0 \, a_-}^2) = 2 \hbar \omega_0 A^2_\mathrm{LO},
\end{equation}
where, as before, $a_+$ and $a_-$ have been included explicitly before
being set to unity.

However, rather than develop an equivalent set of linear algebra
expressions for computing total noise output in the one-photon
formalism, we instead use (\eref{A2}) and (\eref{A2inv}) to define a
one-photon to two-photon conversion matrix
\beq{} \TwoP =
\frac{1}{\sqrt{2}}\smallMatrix{1 & 1 \\ -\I &+\I} \quad \mbox{such
  that} \quad \smallMatrix{a_1 \\ a_2} = \TwoP \smallMatrix{a_+ \\
  a_-^*}. \eeq

The one-photon transfer matrix of any optical system which does not
mix upper and lower audio sidebands (i.e.~any linear system) can then
be expressed in the two-photon formalism as \beq{TfcA} \TnN{} = \TwoP
\cdot \smallMatrix{t_+ & 0 \\ 0 &t_-^*} \cdot \TwoP^{-1}, \eeq where
$t_\pm$ are the transfer coefficients for the upper and lower audio
sidebands.

\begin{acknowledgments}
The authors gratefully acknowledge the support of the National Science
Foundation and the LIGO Laboratory, operating under cooperative
Agreement No. PHY-0757058. This paper has been assigned LIGO Document
No. LIGO-P1400018.
\end{acknowledgments}

\bibliography{bibliographyFile} 

\end{document}